# Source separation as an exercise in logical induction

Kevin H. Knuth

*Center for Advanced Brain Imaging and Department of Cognitive Neuroscience Studies,
The Nathan Kline Institute*

**Abstract.** We examine the relationship between the Bayesian and information-theoretic formulations of source separation algorithms. This work makes use of the relationship between the work of Claude E. Shannon and the "Recent Contributions" by Warren Weaver (Shannon & Weaver 1949) as clarified by Richard T. Cox (1979) and expounded upon by Robert L. Fry (1996) as a duality between a logic of assertions and a logic of questions. Working with the logic of assertions requires the use of probability as a measure of degree of implication. This leads to a Bayesian formulation of the problem. Whereas, working with the logic of questions requires the use of entropy as a measure of the bearing of a question on an issue leading to an information-theoretic formulation of the problem.

## INTRODUCTION

The problem of source separation concerns the identification of source signals given only detected mixtures of those signals along with some prior knowledge of the physical situation under consideration, such as the origin, propagation, and detection of source activity. This problem is very closely related to the problems of source localization (Knuth & Vaughan 1998; Fry & Bierbaum 2000), blind deconvolution (Bretthorst 1992, Bell & Sejnowski 1995), and even waveform interpolation (Bretthorst 1992).

Aside from various cumulant-based algebraic methods (Tucker 1964; Cardoso & Souloumiac 1993; Cardoso 1995; Cardoso & Comon 1996; De Lathauwer et al. 1995), the majority of algorithms relied on the design of a neural network-based architectures, which utilized minimum mutual information (Infomax) (Nadal & Parga 1994; Bell & Sejnowski 1995; Cardoso 1997; Lee & Orglmeister 1997) or relative entropy (Amari 1996; Pearlmutter & Parra 1996; Yang & Amari 1997) to obtain optimal solutions. While these information-theoretic techniques have dominated the scene, early demonstrations that equivalent or similar results could be obtained by using maximum likelihood (Gaeta & Lacoume 1990; Pham et al. 1992; MacKay 1996; Cardoso 1997) have lead to the application of probability theory via Bayesian inference to the problem (Knuth 1998a,b, 1999; Roberts 1998; Rowe 1999; Everson and Roberts 1999; Mohammad-Djafari 1999). The fact that information theory, historically based on the efficiency of signal transmission, and probability theory, based on inferential relationships among assertions, can lead to identical algorithms suggests a deeper relationship between the two fields. Indeed, this relationship between transmission of messages and the semantic meaning of messages was recognized quite early on by Weaver (Shannon & Weaver 1949) and is the main thrust of Weaver's "Recent Additions", although it took Richard Cox to put this relationship into a precise mathematical form.

## THE LOGICAL STRUCTURE OF INFERENCE AND INQUIRY

Richard T. Cox is perhaps best known for having proved that probability is the unique logically-consistent measure of a relative degree of implication among assertions (Cox 1946, 1961). Using Boolean logic, an assertion $a$ implies an assertion $b$, written $a \rightarrow b$, if $a \wedge b = a$ and $a \vee b = b$, where $\wedge$ is the logical *and* operation such that $a \wedge b$ is an assertion that tells what $a$ and $b$ tell jointly, and $\vee$ is the logical *or* operation such that $a \vee b$ is an assertion that tells what $a$ and $b$ tell in common. As an example consider two assertions $a =$ *"It is French bread!"* and $b =$ *"It is food!"*. [1]

Generalizing Boolean implication to implication of varying degree one defines probability as the degree to which the implicate $b$ is implied by the implicant $a$, $p(b|a) \equiv (a \rightarrow b)$, where the right-hand side symbolizes the real value representing the relative degree to which $a$ implies $b$. The inferential utility of this formalism is readily apparent when the implicant is an assertion representing a premise and the implicate is an assertion representing a hypothesis.

Cox (1979) continues by defining a question in terms of an exhaustive set of assertions, which answer the question. Just as inference was examined using the relationships among assertions under the action of implication, one can examine inquiry by considering the bearing that one question may have on another question. The relationships among questions are symmetric to that of assertions, such that a question $A$ has bearing on issue $B$, written $A \rightarrow B$, if $A \wedge B = A$ and $A \vee B = B$, where $\wedge$ is the logical *and* operation such that $A \wedge B$ is a question that asks what $A$ and $B$ ask jointly, and $\vee$ is the logical *or* operation such that $A \vee B$ is a question that asks what $A$ and $B$ ask in common. As an example consider two questions $A =$ *"Is this my coffee?"* and $B =$ *"Is this my drink?"*. The joint question, "Is this my coffee and is this my drink?" asks the same as "Is this my coffee?", whereas the common question, "Is this my coffee or is this my drink?" asks what the two questions ask in common, "Is this my drink?". In this example the question $A$ has bearing on the question $B$.

Generalizing this Boolean relationship to relative degrees of bearing, $b(A|B) \equiv (A \rightarrow B)$, one obtains relationships among questions that are analogous to those among assertions. The bearing of a question on an issue can be written in terms of the entropy using the probabilities of the assertions defining the question and the premise corresponding to the issue. In this sense, probability quantifies what is known, whereas entropy quantifies what is not known.

---

[1] Here we adopt the notation used by Cox where an assertion is denoted by a lowercase Roman character, and a question is denoted by an uppercase Roman character. Subscripts may be used to distinguish components of a compound assertion or question. In addition, we adopt the notation used by Fry where assertions are stated with exclamation marks and questions with question marks. Please note that compound assertions and questions are denoted using boldface block lettering.

## SOURCE SEPARATION AS INDUCTIVE INFERENCE

We begin the application of inductive logic by working in the space of assertions and considering source separation as a problem in inductive inference. To illustrate the relationships between the techniques, we keep the problem simple by assuming linear, stationary, noiseless mixing of $n$ source signals recorded by $n$ detectors. The signal recorded by detector $i$ at time $t$ is given by

$$x_{it} = a_{ij} s_{jt} \qquad (1)$$

where the Einstein summation convention is employed with $s_{jt}$ as the $j$th source signal at time $t$ and $a_{ij}$ as the mixing matrix element describing the coupling between source $j$ and detector $i$. We further assume that the sources are statistically independent and that we possess some prior knowledge regarding the form of the source amplitude densities, denoted $p(s|h)$, where $h$ represents the premise (also known as the prior information).

The model consists of two assertions:
>    **a** =*"The mixing matrix is  a!"*
>    **s** =*"The source waveforms are  s!"*

that are utilized with two additional assertions, one stating the observed data
>    **x** =*"The recorded mixtures are  x!"*

and one stating the premise
>    **h** =*"Our prior state of knowledge is ... !"*.

Note that the assertions are all compound assertions in the sense that they are conjunctions of more simple assertions. For example, the assertion **a** regarding the mixing matrix is an assertion composed of the conjunction of $n^2$ assertions regarding the coupling terms between the sources and the detectors, such that

$$\mathbf{a} = a_{11} \wedge a_{12} \wedge \ldots \wedge a_{21} \wedge \ldots \wedge a_{2n} \wedge \ldots \wedge a_{nn}. \qquad (2)$$

Similarly, **s** and **x** are compound assertions each being the conjunction of $nT$ assertions, one for each source or detector and for each of the $T$ measurements.

Bayes' Theorem is used to express the joint posterior probability of the model of the physical process in terms of the likelihood of the data and the prior probabilities of the model parameters

$$p(\mathbf{a} \wedge \mathbf{s} | \mathbf{x} \wedge \mathbf{h}) = p(\mathbf{a} \wedge \mathbf{s} | \mathbf{h}) \frac{p(\mathbf{x} | \mathbf{a} \wedge \mathbf{s} \wedge \mathbf{h})}{p(\mathbf{x} | \mathbf{h})}. \qquad (3)$$

In this problem, we are interested in finding the inverse of the mixing matrix, the separation matrix **w**. If we are not explicitly interested in the source waveforms, one may obtain the posterior for the mixing matrix alone by marginalizing over the source data

$$p(\mathbf{a} | \mathbf{x} \wedge \mathbf{h}) = \frac{p(\mathbf{a}|\mathbf{h})}{p(\mathbf{x}|\mathbf{h})} \int d\mathbf{s}\, p(\mathbf{s}|\mathbf{h}) p(\mathbf{x}|\mathbf{a} \wedge \mathbf{s} \wedge \mathbf{h}). \qquad (4)$$

The assumption of noiseless mixing implies that one should assign a delta function likelihood for the measurements at a single time $t$.

$$p(\mathbf{x_t} | \mathbf{a} \wedge \mathbf{s_t} \wedge \mathbf{h}) = \prod_i \delta(x_{it} - a_{ij} s_{jt}), \qquad (5)$$

which allows one to readily solve the integral obtaining (for one measurement at time $t$)

$$p(\mathbf{a}|\mathbf{x_t} \wedge \mathbf{h}) = \frac{p(\mathbf{a}|\mathbf{h})}{p(\mathbf{x_t}|\mathbf{h})} \frac{1}{\det \mathbf{a}} \prod_l p_l \left(a_{lk}^{-1} x_{kt}\right), \qquad (6)$$

where the independence of the source waveforms has been used to factorize their joint probability.

For simplicity, we assume that nothing is known about the mixing process and thus assign a uniform prior to $p(\mathbf{a}|\mathbf{h})$. If one is interested only in the maximum of the posterior, it is beneficial to work with the logarithm of the probability

$$\log p(\mathbf{a}|\mathbf{x_t} \wedge \mathbf{h}) = -\log \det \mathbf{a} + \log \prod_l p_l \left(a_{lk}^{-1} x_{kt}\right) + C \qquad (7)$$

where the constant $C$ absorbs the logarithm of the uniform mixing matrix prior and the evidence. Considering data at any time point, this posterior can be maximized with respect to the separation matrix using a stochastic gradient ascent update rule by considering the derivative of the posterior with respect to the separation matrix elements

$$\frac{\partial}{\partial w_{ij}} \log p(\mathbf{a}|\mathbf{x} \wedge \mathbf{h}) = a_{ji} + x_j \left( \frac{p'_i \left(a_{ik}^{-1} x_{kt}\right)}{p_i \left(a_{ik}^{-1} x_{kt}\right)} \right)_j \qquad (8)$$

resulting in the update rule

$$\triangle w_{ij} \propto a_{ji} + x_j \left( \frac{p'_i \left(a_{ik}^{-1} x_{kt}\right)}{p_i \left(a_{ik}^{-1} x_{kt}\right)} \right)_j \qquad (9)$$

This equation, derived in more detail elsewhere (Knuth 1998a,b; 1999), is identical to the results obtained by Bell & Sejnowski (1995), however it has been derived from the viewpoint of inductive inference by working in the space of assertions.

## SOURCE SEPARATION AS INDUCTIVE INQUIRY

The problem of source separation is now looked at from the viewpoint of inductive inquiry by working in the space of questions. Dual to the premise **h** is the issue "*What are the characteristics of this mixing problem?*", which we shall denote by **H**. The questions we will consider are

$\mathbf{S} = $ "*What are the original source signals?*"
$\mathbf{M} = $ "*What is the mixing matrix?*"
$\mathbf{X} = $ "*What are the recorded signals?*"

Note again that all of these questions are compound questions, where **X** asks $\mathbf{X} = X_1 \wedge X_2 \wedge \ldots \wedge X_n$ with $X_1 = $ "What is the signal recorded in detector 1?" and similarly for **S** and **M**. The question $\mathbf{S} \wedge \mathbf{M}$ equivalently asks "What is the physical description of this source separation problem?".

Unfortunately, the question $\mathbf{S} \wedge \mathbf{M}$ cannot be asked directly, so one must model the source separation problem and ask questions like

$\mathbf{Y} =$ *"How well have we modeled the source activity?"*

To do this one considers what the question $\mathbf{S} \wedge \mathbf{M}$ has in common with $\mathbf{Y}$. The questions $\mathbf{S}$, $\mathbf{M}$ and $\mathbf{Y}$ are related such that if one answers question $\mathbf{S} \wedge \mathbf{M}$, one has answered question $\mathbf{Y}$. Thus $\mathbf{S} \wedge \mathbf{M}$ implies $\mathbf{Y}$, and from the definition of implication

$$((\mathbf{S} \wedge \mathbf{M}) \vee \mathbf{Y}) = \mathbf{Y}. \tag{10}$$

We consider what questions $\mathbf{Y}$ and $\mathbf{X}$ ask in common and look at the bearing of the common question $\mathbf{Y} \vee \mathbf{X}$ on the issue $\mathbf{H}$, $b(\mathbf{Y} \vee \mathbf{X}|\mathbf{H})$

$$b(\mathbf{Y} \vee \mathbf{X}|\mathbf{H}) = b(\mathbf{Y}|\mathbf{H}) - b(\mathbf{Y} \vee \sim \mathbf{X}|\mathbf{H}). \tag{11}$$

By translating the bearing notation to conventional information-theoretic notation one finds that this is equivalent to Bell and Sejnowski's (1995) eqn 2.1

$$I(Y;X) = H(Y) - H(Y|X) \tag{12}$$

where the mutual information, $I(Y;X)$, is the bearing of the common question $\mathbf{Y} \vee \mathbf{X}$ on the issue $\mathbf{H}$ and the conditional entropy, $H(Y|X)$, is equivalent to the bearing of the question $\mathbf{Y} \vee \sim \mathbf{X}$ on the issue $\mathbf{H}$.

We now consider how one should vary the modeled mixing matrix, or the separation matrix, so that the bearing of the common inquiry on the source separation issue can be maximized. Since the common question "From what has not been recorded by the detectors, have the source signals have been overestimated?", $\mathbf{Y} \vee \sim \mathbf{X}$, refers to the source signal components never recorded by the detectors, this inquiry does not depend on the mixing process, which describes how the source signals are mapped onto the detectors. So the variation of the bearing of this inquiry on the issue with respect to the elements of the separation matrix, $w_{ij}$, is identically zero

$$\frac{\partial b(\mathbf{Y} \vee \sim \mathbf{X}|\mathbf{H})}{\partial w_{ij}} \equiv 0. \tag{13}$$

So that maximizing the bearing of the common inquiry on the issue, $b(\mathbf{Y} \vee \mathbf{X}|\mathbf{H})$, with respect to the elements of the separation matrix one requires

$$\frac{\partial b(\mathbf{Y}|\mathbf{H})}{\partial w_{ij}} = 0, \tag{14}$$

so that we can concentrate on the bearing of the inquiry of the quality of the model on the issue.

By writing the bearing in terms of entropy calculated using the premise that corresponds to the issue, one obtains

$$\frac{\partial b(\mathbf{Y} \vee \mathbf{X}|\mathbf{H})}{\partial w_{ij}} = -\frac{\partial}{\partial w_{ij}} \int \mathbf{dy}\, p(\mathbf{y}|\mathbf{h}) \log p(\mathbf{y}|\mathbf{h}), \tag{15}$$

which when set to zero allows one to locate the separation matrix that maximizes this bearing.

The assertion **y** is a compound assertion consisting of the conjunction of $n$ assertions $\mathbf{y} = y_1 \wedge y_2 \wedge \ldots \wedge y_n$ where $y_i$ is the degree to which the source waveforms may have been overestimated. This can be addressed by transforming an estimate of the source amplitude $s_{it} = a_{ik}^{-1} x_{kt}$ by the prior probability *distribution* of the source amplitudes

$$y_{it} = q(s_{it}|\mathbf{h}) \equiv \int_{-\infty}^{s_{it}} ds\, p(s|\mathbf{h}), \tag{16}$$

where $p(s|\mathbf{h})$ is the probability *density* of the source amplitude. A severe underestimate will result in a value approaching 0, whereas a severe overestimate will result in a value approaching 1. Using the fact that the mixing is noiseless, we can write the source waveforms as $\mathbf{s} = \mathbf{a}^{-1}\mathbf{x}$ and perform a change of variables in Eqn. 15. using

$$p(y|h) = p(x|h) \frac{|dy|}{|dx|}^{-1} \tag{17}$$

to find, again for a single measurement at time $t$,

$$\frac{\partial b(\mathbf{Y} \vee \mathbf{X}|\mathbf{H})}{\partial w_{ij}} = -\frac{\partial}{\partial w_{ij}} \int d\mathbf{x}\, p(\mathbf{x}|\mathbf{h}) \log \left| \left( \det \mathbf{a}^{-1} \prod_l p_l \left(a_{lk}^{-1} x_{kt}\right) \right)^{-1} p(\mathbf{x}|\mathbf{h}) \right|. \tag{18}$$

The argument of the logarithm breaks into the sum of logarithms of three terms, of which the term containing $\log p(\mathbf{x}|\mathbf{h})$ yields zero when the derivative with respect to the mixing matrix is applied. Again we can implement a stochastic gradient ascent method by utilizing the nonzero terms of the argument of the expectation value above

$$\triangle w_{ij} = \frac{\partial}{\partial w_{ij}} \left( -\log \det \mathbf{a} + \log \prod_l p_l \left(a_{lk}^{-1} x_{kt}\right) \right), \tag{19}$$

which when evaluated gives results identical to those found by working in the space of assertions (Eqn. 9). Under this inductive logic framework, one can see precisely how the results obtained by Bell & Sejnowski (1995) using information theory are related to the results obtained by Knuth (1998a, b, 1999) using Bayesian inference.

## CONCLUSIONS

Shannon's work dealt with a simplified case of inductive inquiry where one has a transmitter and is attempting to design a communication channel from the transmitter to the receiver, equivalently designing a question to be answered by the transmitter. This problem, which dealt with at most two questions, led to the development of information theory. Even though the calculus of inductive inquiry in simple cases is identical to information theory, the logical foundation of the inferential inquiry approach provides a notation that is based on the logical relations among questions. This leads to relations that are frequently conceptually more intuitive than their information-theoretic counterparts. As an example, where Bell and Sejnowski attribute the conditional entropy in their equation to the entropy of the noise, we found that it actually represents the effect that information about the sources not present in the data has on the inquiry. In addition, there are relationships that can be derived using the inferential calculus that have no information-theoretic counterpart. As an example, consider the bearing of the common question $A \vee B \vee C$ on an issue $H$, which cannot be dealt with in an information-theoretic framework.

Weaver's "Recent Contributions" have often gone relatively unnoticed perhaps due to the fact that he was unable to quantify his impressions of the relationship between information theory as derived by Shannon and semantic meaning. This interrelation having been quantified by Cox as a duality between the space of assertions (being appropriate for performing inductive inference and hence determining semantic meaning) and the space of questions (being appropriate for performing inductive inquiry and hence related to the choosing of a channel or question for communication) has now been placed in a theoretical framework that can not only be utilized to understand the relationship between Bayesian and information-theoretic solutions to solved problems, but also to approach a host of new and heretofore inaccessible problems. Robert Fry's work on inductive logic has greatly expounded on Cox's initial quantification and has demonstrated the logical foundations of neural network processing and have expanded on their application (Fry 1995, 1996, 2000).

By now, you will have probably realized that there are several different bearings that one may wish to maximize, just as there may be different posterior probabilities to be evaluated. We have looked at one particular question regarding the quality of our source separation model and found that this reproduces precisely the approach taken by Bell and Sejnowski using a neural network architecture in conjunction with a minimum mutual information or infomax principle. The connection between the inferential inquiry approach and the information-theoretic approach sheds some light on the logical basis of Bell and Sejnowski's work and on neural network design in general. In addition, the duality between the logic of assertions and the logic of questions elucidates the relationship between Bayesian and information-theoretic approaches. However, inductive inquiry goes far beyond information theory and provides a means to work with the logic of questions directly. This way of thinking is quite different than that required by the logic of assertions. To use this new logic effectively will require a paradigm shift in our thinking about problems. To begin with, there is a need to determine better questions to answer.

## ACKNOWLEDGEMENTS

I would like to thank Robert Fry for taking the time and effort to introduce me to this fascinating work, Larry Bretthorst for many insightful discussions, and Daniel Javitt and Robert Bilder for their continued support.

**References**

Amari, S. 1996. Natural gradient works efficiently in learning. *Neural Comp* 10:251-276.

Bell A.J. and Sejnowski T.J. 1995. An information-maximization approach to blind source separation and deconvolution, *Neural Comp*, 7:1129-1159.

Bretthorst G.L. 1992. Bayesian interpolation and deconvolution. Technical Report CR-RD-AS-92-4, The U. S. Army Missile Command

Cardoso, J.-F. 1995. A tetradic decomposition of 4th -order tensors: application to the source separation problem. In: *SVD and signal processing III: algorithms, applications and architectures*, B. De Moor, M. Moonen (eds.).. Elsevier Science Publishers, North Holland, Amsterdam, pp. 375-382.

Cardoso, J.-F. 1997. Infomax and maximum likelihood for source separation, *IEEE Letters on Signal Processing*, 4(4):112-114.

Cardoso, J.-F. and Souloumiac, A. 1993. An efficient technique for the blind separation of complex sources. In: *Proceedings IEEE Signal Processing Workshop on Higher-Order Statistics*, Lake Tahoe, USA, pp. 275-279.

Cardoso, J.-F. and Comon, P. 1996. Independent component analysis, a survey of some algebraic methods. In: *Proceedings of IEEE International Symposium on Circuits and Systems*, 2:93-96.

Cox R.T. 1946. Probability, frequency, and reasonable expectation, *Am. J. Physics*, 14:1-13.

Cox R.T. 1961. *The Algebra of Probable Inference*, The Johns Hopkins Press, Baltimore.

Cox R.T. 1979. Of inference and inquiry, In *Proc. 1978 Maximum Entropy Formalism Conference*, MIT Press, pp.119-167.

De Lathauwer, L., De Moor, B., Vandewalle, J. 1995. The application of higher order singular value decomposition to independent component analysis. In: *SVD and signal processing III: algorithms, applications and architectures*, B. De Moor, M. Moonen (eds.), Elsevier Science Publishers, North Holland, Amsterdam.

Everson R. and Roberts S.J. 1999. Independent component analysis: a flexible non-linearity and decorrelating manifold approach. *Neural Comp*, 11(8).

Fry R.L. 1995. Observer-participant models of neural processing, *IEEE Trans. on Neural Networks*, 6(4):918-928.

Fry R.L. 1996. Rational neural models based on information theory, in K.M. Hanson and R.N. Silver (eds.), *Maximum Entropy and Bayesian Methods, Santa Fe 1996*, Dordrecht: Kluwer Academic Publishers, pp. 335-340.

Fry R.L. 2000. Cybernetic systems based on inductive logic, in A. Mohammad-Djafari (ed.), *Maximum Entropy and Bayesian Methods, Paris 2000*, Dordrecht: Kluwer Academic Publishers.

Fry R.L. and Bierbaum M.M. 2000. Bayesian source separation application to sensor handover problem, unpublished.

Gaeta, M. and Lacoume, J.-L. 1990. Source separation without prior knowledge: the maximum likelihood solution. In: *Proceedings of the European Signal Processing Conference (EUSIPCO'90)*, pp. 621-624.

Knuth K.H. 1998a. Difficulties applying recent blind source separation techniques to EEG and MEG. In *Maximum Entropy and Bayesian Methods, Boise 1997*, G. J. Erickson, J.T. Rychert and C.R. Smith (eds.), Dordrecht: Kluwer Academic Publishers, pp. 209-222.

Knuth K.H. 1998b. Bayesian source separation and localization, in *Proceedings of SPIE: Bayesian Inference for Inverse Problems*, vol. 3459, A. Mohammad-Djafari (ed.), pp. 147-158.

Knuth K.H. 1999. A Bayesian approach to source separation. In *Proceedings of the First International Workshop on Independent Component Analysis and Signal Separation: ICA'99*, J.-F. Cardoso, C. Jutten and P. Loubaton (eds.), , Aussios, France, Jan. 1999, pp. 283-288.

Knuth K.H. and Vaughan H.G., Jr. 1999. The Bayesian origin of blind source separation and electromagnetic source estimation. In *Maximum Entropy and Bayesian Methods, Munich 1998*, W. von der Linden, V. Dose, R. Fischer, and R. Preuss (eds.), Dordrecht: Kluwer Academic Publishers., pp. 217-226.

Lee, T.-W. and Orglmeister, R. 1997. A contextual blind separation of delayed and convolved sources. In: *Proceedings of the IEEE International Conference on Acoustics, Speech and Signal Processing*, Munich, pp. 1199-1203.

MacKay D.J.C. 1996. Maximum likelihood and covariant algorithms for independent component analysis, Draft Paper: http://wol.ra.phy.cam.ac.uk/mackay/

Mohammad-Djafari A. 1999. A Bayesian approach to source separation. In *Maximum Entropy and Bayesian Methods, Boise 1999*, G. J. Erickson, J.T. Rychert and C.R. Smith (eds.), Dordrecht: Kluwer Academic Publishers.

Nadal, J.-P. and Parga, N. 1994. Non-linear neurons in the low-noise limit: a factorial code maximizes information transfer. *Network*, 4:295-312.

Pearlmutter, B.A. and Parra, L.C. 1996. A context-sensitive generalization of ICA, *1996 International Conference on Neural Information Processing*, Hong Kong.

Pham D.T. Garrat P and Jutten C. 1992. Separation of a mixture of independent sources through a maximum likelihood approach, in *Proc. EUSIPCO*, p.771-774.

Roberts S.J. 1998. Independent component analysis: source assessment & separation, a Bayesian approach. *IEE Proceedings - Vision, Image & Signal Processing*, 145(3), 149-154.

Rowe, D.B. 1999. A Bayesian Approach to Blind Source Separation. Preprint.

Shannon C.E. & Weaver W. 1949. *The Mathematical Theory of Communication*, The University of Illinois Press, Urbana.

Tucker, L.R. 1964 The extension of factor analysis to three-dimensional matrices. In *Contributions to mathematical psychology*, H. Gulliksen, N.

Frederiksen (eds.).. Holt, Rinehart & Winston, pp. 109-127.

Yang, H.H. and Amari S. 1997. Adaptive on-line learning algorithms for blind separation - maximum entropy and minimum mutual information, *Neural Comp*, 9(7):1457-1482.